# Optical vortex-induced forward mass transfer: Manifestation of helical trajectory of optical vortex

RYOSUKE NAKAMURA[1], HARUKI KAWAGUCHI[1], MUNEAKI IWATA[2], AKIHIRO KANEKO[3], RYO NAGURA[4], SATOYUKI KAWANO[4], KOHEI TOYODA[1,5], KATSUHIKO MIYAMOTO[1,5], AND TAKASHIGE OMATSU[1,5,*]

[1] *Graduate School of Engineering Chiba University, Chiba 263-8522, Japan*
[2] *Ricoh Institute of Information and Communication Technology, Applied Imaging Development Center, Kanagawa 243-0460, Japan*
[3] *Ricoh CT&P Division.1st Technology Center, Shizuoka 410-0007, Japan*
[4] *Graduate School of Engineering Science, Osaka University, Japan*
[5] *Molecular Chirality Research Center, Chiba University, Chiba 263-8522, Japan*

*\*omatsu@faculty.chiba-u.jp*

**Abstract:** The orbital angular momentum of an optical vortex field is found to twist high viscosity donor material to form a micron-scale 'spin jet'. This unique phenomenon manifests the helical trajectory of the optical vortex. Going beyond both the conventional ink jet and laser induced forward mass transfer (LIFT) patterning technologies, it also offers the formation and ejection of a micron-scale 'spin jet' of the donor material even with an ultrahigh viscosity of 4 Pa·s. This optical vortex laser induced forward mass transfer (OV-LIFT) patterning technique will enable the development of next generation printed photonic/electric/spintronic circuits formed of ultrahigh viscosity donor dots containing functional nanoparticles, such as quantum dots, metallic particles and magnetic ferrite particles, with ultrahigh spatial resolution. It can also potentially explore a completely new needleless drug injection.

## 1. Introduction

The nozzle based ink jet technique is capable of printing anywhere with selectivity using a micrometer-scale liquid droplet to shape various patterns; therefore, it is widely used in fields such as color image printing, printed (or flexible) photonics/electronics/spintronics and integrated optical circuits as a non-contact process [1–3]. However, this technique has several drawbacks, in that it is difficult to form and eject high viscosity droplets containing functional nanoparticles, such as quantum dots, metallic particles and magnetic ferrite particles, with viscosities greater than 0.1 Pa·s. [4] In addition, the smallest jet droplet diameter is typically about 20 μm.

Laser induced forward mass transfer (LIFT) [5–7], a direct deposition approach without any nozzle clogging, enables the selective transfer of the donor material towards a receiver substrate with high spatial resolution. To date, the patterning of various donor materials, such as metals and biological elements with liquid or solid phases based on LIFT have been demonstrated. However, it is still difficult to efficiently pattern micrometer-scale dots of ultrahigh viscosity donor materials on a receiver substrate separated from the donor material layer by a millimeter-scale distance.

An optical vortex [8–14] has a doughnut shaped spatial form and an orbital angular momentum (OAM) that arises from its helical wavefront with a phase singularity characterized by a topological charge, $\ell$. Circularly polarized light also carries a spin angular momentum (SAM) associated with its helical electric field [15]. In recent years, the authors and co-workers reported that a laser materials processing technique, employing an optical field with OAM, enables the fabrication of unique material structures, such as chiral metal structures and silicon needles on an irradiated target assisted by the SAM [16–20]. During fabrication of silicon needles, the optical vortex was determined to provide a spin on molten silicon droplets (viscous droplets), which resulted in the efficient accumulation and straight flight of micrometer-scale molten silicon droplets, so-called 'silicon jet'. [21] Such an optical vortex should provide an entirely new technique for applications such as the patterning of ultrahigh viscosity donor droplets with ultrahigh spatial resolution and extremely long flight distance, beyond both conventional nozzle based ink jet and LIFT technologies.

Here, we report for the first time the formation of a micrometer-scale 'spin jet', made of ultrahigh viscosity donor materials, using optical vortex laser induced forward mass transfer (OV-LIFT). The optical vortex forces the molten donor material to axially spin, thereby creating efficiently a 'spin jet'. This phenomenon manifests the helical trajectory of the optical vortex, and it will be extended to provide entirely novel insights into fundamental and advanced sciences, for instance, new generation LIFT techniques, enabling forward mass transport of ultrahigh viscosity donor materials with ultrahigh spatial resolution and extremely long flight distance (millimeter scale). Ultrahigh viscosity materials possibly decelerate the OAM induced spinning motion of the jet and droplets. Thus, we also address the direct observation of the interaction between the OAM field and the donor materials.

2. **Experimental setup**

Fig. 1 shows a schematic illustration of the experimental setup for OV-LIFT. A nanosecond green laser (frequency-doubled Q-switched Nd:YAG laser) with a wavelength of 532 nm, a maximum pulse energy of ~100 µJ and a pulse duration of ~3 ns was used, and its output was converted to a circularly polarized first-order optical vortex with $\ell=1$ and $s=1$ by employing a spiral phase plate (SPP) [22] and a quarter-wave plate (QWP). A donor was an air-dried pigment film with a thickness of about 20 µm, formed of an ink diluted with ethyl acetate as a solvent, on a silica glass substrate. The film exhibited a viscosity of about 4 Pa·s (this value is approximately 1000 times higher than that for water and 100 times higher than that of ink used for conventional ink jet printing). It also exhibited optical density of >5 at 532 nm, and thermal conductivity of 0.35 $Wm^{-1}K^{-1}$.

The 532 nm optical vortex pulse (single vortex pulse deposition) was loosely focused to form a 160 µm diameter annular spot on the donor film from the backside (the silica glass plate substrate side). Temporal evolution of the jet ejected from the donor film was observed with a high speed camera (Shimadzu Corp., Hyper Vision HPV-X) from the side at a frame rate of $2\times10^6$ fps.

3. **Results and discussion**

The irradiated donor film was deformed and underwent radial inward mass transport towards the dark core of the optical vortex. After irradiation by the optical vortex pulse, the donor film formed a jet within approximately 4 µs (Fig. 2(a), see supplementary file 1). About 6 µs later, a micrometer-scale single droplet (with a diameter of ~20 µm) was ejected from the tip of the jet due to Plateau-Rayleigh instability [23–25], and formed a circular dot with a diameter of about 22 µm on glossy paper set at least 1 mm away from the donor film (Fig. 2(b)). The pulse energy was then measured to be ~47 µJ. It should be noted that both the jet and single droplet were formed within a pulse energy range of 38-52 µJ. When the energy was lower than this level, droplets were not ejected. At a higher energy level, the donor film was decomposed into multiple fragments that were ejected together. It is worth noting that the pulse energy required for the formation of the jet will be dependent on the film thickness and the viscosity of the donor material itself. The axial velocity of the front edge of the jet (we call it 'the flight speed of the jet') and the flight speed of a single droplet were well fitted by a linear function of the pulse energy, as shown in Fig. 3(a), and they were measured to be 10-80 m/s. The product of $kR_0$, where $k$ is the wavenumber ($=2\pi/\lambda_{ink}$; $\lambda_{ink}$ is the wavelength of instability) of the jet along the propagation direction and $R_0$ is the unperturbed inner radius of the jet as shown in Fig. 2(c), was estimated to be less than unity (approximately 0.26~0.65) within the pulse energy range of 38-52 µJ, manifesting Plateau-Rayleigh instability (Fig. 3(b)). It is also worth noting that the experimental $kR_0$ at the pulse energy of >50 µJ is very close to 0.697 obtained by the linear instability theory, manifesting the fastest growing instability. Further, the $\lambda_{ink}$ and $R_0$ were then measured to be 54-121 µm and 5-6 µm, respectively (Fig. 3(c)). Here, assuming that the jet undergoes Plateau–Rayleigh instability and it collapses into a spherical droplet, the diameter of the droplet can be estimated to be 22~28 µm from its volume given by $\pi R_0^2 \lambda_{ink}$. This value also supports well the experimental one (~22 µm) as shown in Fig. 2(b).

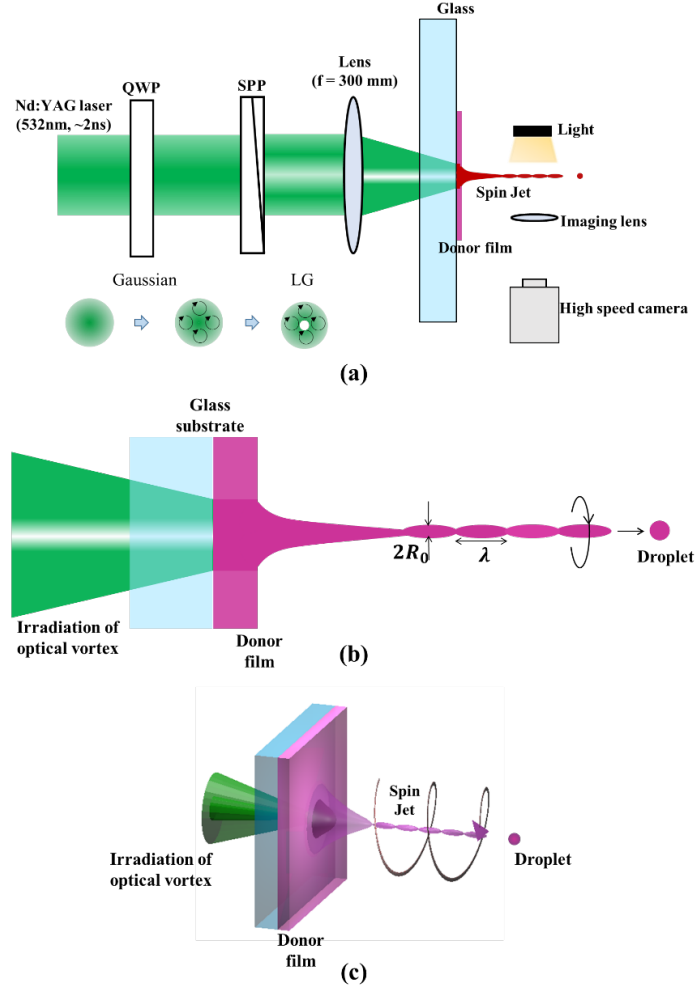

Fig. 1. Experimental setup for OV-LIFT. The pump laser used was a nanosecond green laser with a wavelength of 532 nm and a pulse width of ~3 ns. A spiral phase plate and a quarter-wave plate were also employed to convert the pump laser output to the circularly polarized optical vortex. All experiments were performed at room temperature and in ambient air. (b) Magnified 'spin jet'. $R_0$ is the inner radius of the jet, and $\lambda_{ink}$ is the wavelength of the jet along the propagation direction. (c) 3-D image of OV-LIFT induced 'spin jet'.

In contrast, irradiation with even a conventional circularly polarized Gaussian pulse, *i.e.*, an optical field with SAM and without OAM, decomposed the donor film into many micrometer-scale fragments, without formation of a jet, at any pulse energies (Fig. 2(d), see supplementary file 2).

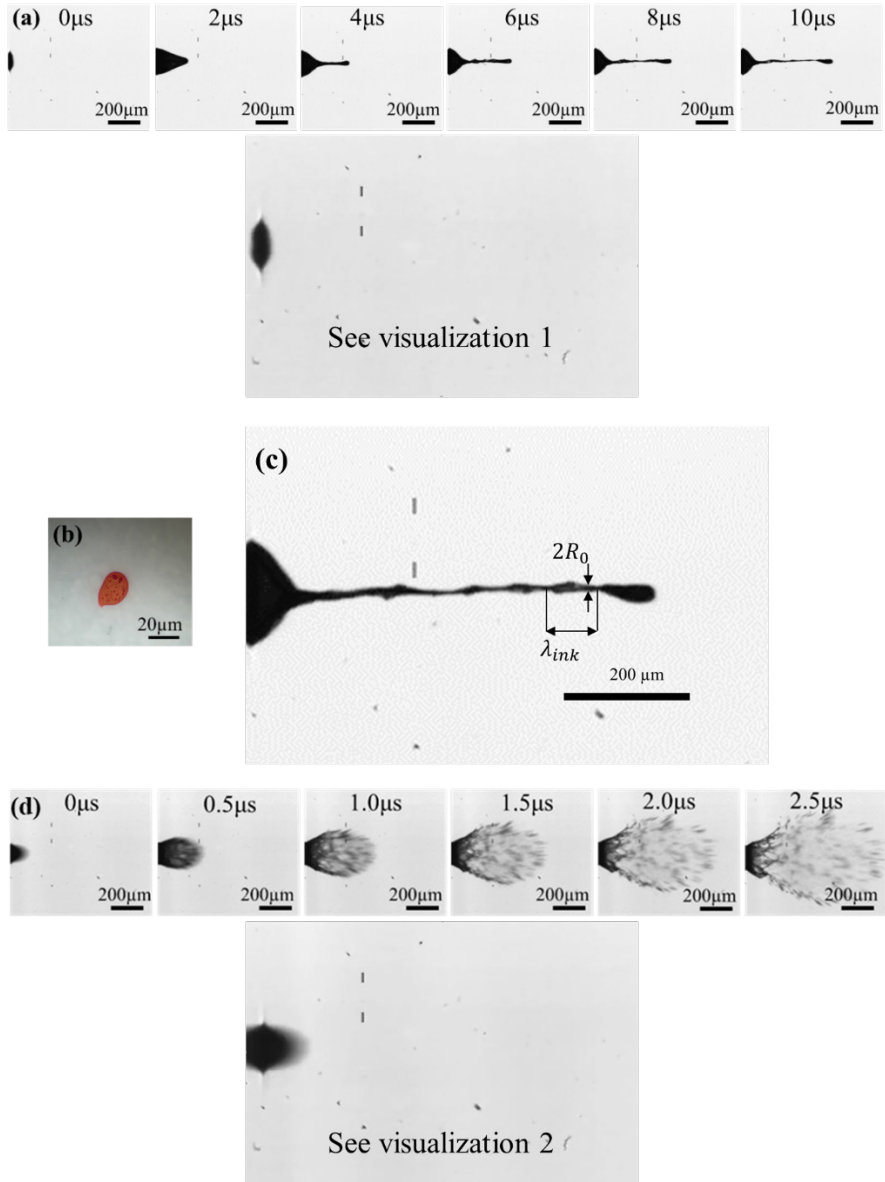

Fig. 2. (a) Time lapse for the formation of a jet by irradiation of optical vortex. The optical vortex pulse carried an OAM with $\ell=1$ and a SAM with $s=1$ (see Visualization 1). (b) Ink dot (with a diameter of ~22 µm) printed on glossy paper 1 mm away from the film. (c) Time lapse of the jet at 10 µs after laser irradiation. The unperturbed inner radius of the jet, $R_0$, and the wavelength of instability of jet, $\lambda_{ink}$, are measured. (d) Time lapse of scattered droplets formed by irradiation with a circularly polarized Gaussian beam. Both optical vortex and Gaussian pulse energies were measured to be 47 µJ (see Visualization 2).

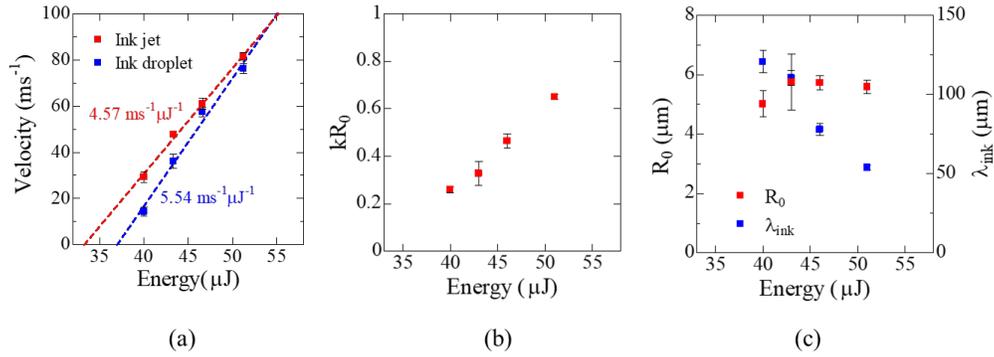

Fig. 3. (a) Experimental flight speeds of the jet and single droplet as a function of the irradiated pulse energy. (b) The product of $kR_0$, *i.e.* figure of merit for Plateau-Rayleigh instability at various pulse energies. (c) The unperturbed inner radius of the jet, $R_0$, and the wavelength of instability of jet, $\lambda_{ink}$, at various pulse energies.

Why does irradiation with an optical vortex pulse allow the formation of such a jet? The irradiated donor film is considered to be molten and thermally deformed. Irradiation with a ring-shaped optical vortex induces a non-uniform vaporization pressure [26–28] and shockwave that collects the molten donor material within its dark core, with the help of the optical radiation forward pressure, which results in the formation of a jet. It should also be noted that the high viscosity of the donor material assists the formation of the jet. The superfluid liquid-phase molten donor material is also pinched off to form the shape of a spherical droplet due to surface tension. The OAM should then impart a spin to the molten donor material, which assists the formation of the jet and the straight flight of the spherical droplet. In fact, the spinning motion of the droplets was directly observed, as shown in Fig. 4 (see supplementary file 3). The spin rate of the droplet was measured to be about $10^5$ rps. The spin direction of the droplet was also reversed by inversion of the sign of the OAM. Thus, the jet generated by the irradiation of optical vortex is here called a 'spin jet'.

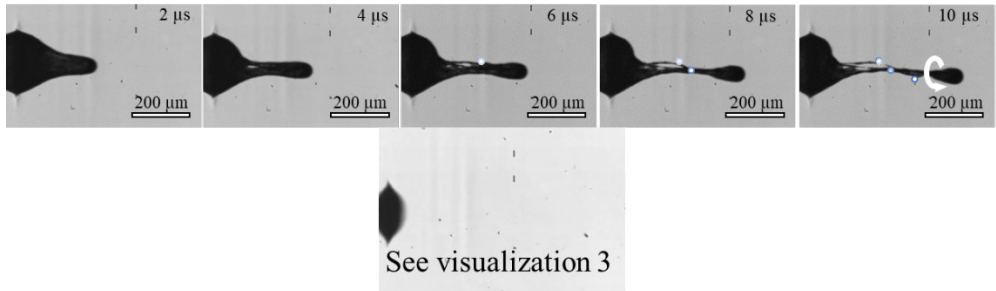

Fig. 4. Time laps of 'spin jets' created by irradiation with right-handed optical vortex pulses. The OAM provides a spin on the molten donor material to create the 'spin jet', as evidenced by spiraling of the jet towards the clockwise direction. Arrows indicate the rotational direction of the 'spin jet'. (see Visualization 3)

How do the OAM and SAM contribute to the formation of the 'spin jet'? The spin rate for the droplet was measured for various total angular momentum (TAM) indices, $J(= -3$ to $3)$, defined as the sum of $\ell$ and $s$. Higher-order vortex modes with a larger topological charge, $\ell$, generally exhibit an expanded spatial mode field with a large dark core. The diameter of the focused vortex modes was then fixed at about 160 μm so as to maintain the intensity of the focused spot on the donor film by optimizing the effective numerical aperture of the focusing lens.

The pulse energy was also fixed to be 47 µJ. The spin direction of the droplet was fully determined by the sign of the OAM, $\ell$, and it was not affected by the SAM, $s$. The spin rate for the droplet increased with the magnitude of $\ell$. The SAM contributes to accelerate or decelerate the spinning motion of a droplet when the product of $\ell$ and $s$ is positive or negative, i.e. $\ell$ has the same sign as or the opposite sign to that of $s$. Also, it is noteworthy that the spin rate of the droplet is determined by the magnitude of TAM. Thus, the SAM assists the stable straight flight of the 'spin jet'. Such degeneracy among optical vortices with the same TAM can be well supported by the previously reported experimental results concerning spiral metal structures formation [18].

The maximum spin rate was about $1.3 \times 10^5$ rpm for $|J|=3$. Further, even optical vortices with $J=0$ ($\ell=\pm 1$; $s=\mp 1$) induced spinning of the droplet, so as to form a jet (Fig. 5).

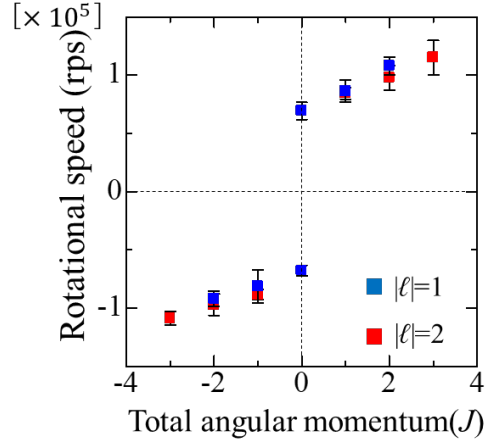

Fig. 5. Measured rotational speed of 'spin jet' as a function of the TAM of the optical vortex field.

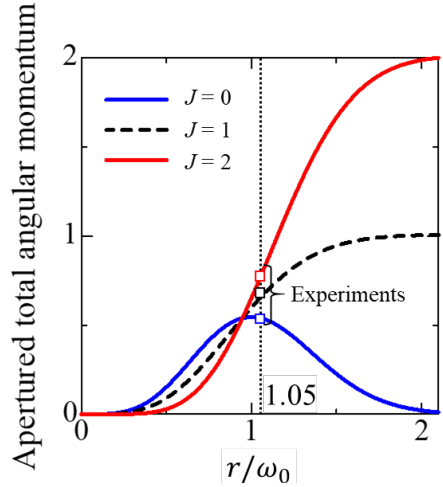

Fig. 6. Apertured total angular momentum at various $J$. The effective TAM aperture was estimated to be about 1.05 $\omega_0$ by assuming that the rotational speed ratio $R_{speed}$ (= 1.0 : 0.88 : 0.69) among $J$=2, 1, and 0 was dominated only by the ratio $R_J$ among $\tilde{J}$(1, 1), $\tilde{J}$(1, 0), and $\tilde{J}$(1, -1), i.e. the effective TAM aperture was determined by the relationship of $R_{speed} = R_J$. The ratio among the vertical values at square plots corresponds to the experimental rotation speed ratio among $J$=2, 1, and 0.

These results can be understood by employing the finitely-apertured TAM [18], $\tilde{J}(r,\ell,s)$, defined as:

$$\tilde{J}(r,\ell,s) = 2\pi \int_0^r j_{\ell,s}(r) r\, dr, \quad (1)$$

$$j_{\ell,s}(r) \propto \frac{1}{\ell!}\left(\ell - |\ell|s + s\left(\frac{\sqrt{2}r}{\omega_0}\right)^2\right)\left(\frac{\sqrt{2}r}{\omega_0}\right)^{2|\ell|} e^{\frac{-2r^2}{\omega_0^2}} \quad (2)$$

where $j_{\ell,s}$ is the total angular momentum density of the optical vortex field with $\ell$ and $s$, and $r$ is the radial axis, as shown in Fig. 6. The relationship $\tilde{J}(r,\ell,s) > \tilde{J}(r,\ell,0) > \tilde{J}(r,\ell,-1)$ is then established at large apertures with $r < \omega_0$. Even an optical vortex mode with zero TAM spatially averaged over the entire beam aperture exhibits non-zero finitely-apertured TAM, which thus imparts spinning motion to the molten donor material. For instance, in the case of $\ell = 1$, the effective TAM aperture was estimated to be about 1.05 $\omega_0$ by assuming that the rotational speed was dominated by the finitely-apertured TAM.

Assuming that the droplet is a rotating sphere in a high viscous medium with a low Reynolds number, the Stokes drag torque, $\tau$, can be estimated using the following expression [29]:

$$\tau = 8\pi\sigma a^3 (2\pi f), \quad (3)$$

where $\sigma$ is the viscosity of the surrounding medium, $a$ is the radius of the sphere, and $f$ is the rotational speed. The OAM field should provide effective torque on the sphere, given by:

$$\tau_{OAM} = \frac{E}{t} J \cdot \frac{\tilde{J}(1.05\omega_0)}{\tilde{J}(\infty)}, \quad (4)$$

where $E$ is the pulse energy and $t$ is the pulse width. Thus, the experimental value (= $1.1\times10^5$ rps) is 0.73 times the rotational speed $f$ (= $1.5\times10^5$ rps) theoretically estimated (this value should be the maximum value) by substituting the values $\sigma = 1.8\times10^{-5}$ (air) Pa·s [30], $a = 22$ μm, $E = 47$ μJ, $t = 3$ ns, and $J = 2$ ($\ell = 1$, $s = 1$) into Eqs. (3) and (4).

The 'spin jet' created by irradiation with a 2$^{nd}$-order optical vortex field with $\ell = 2$ was split into helical pigtails, which were recombined to form a spherical droplet due to surface tension of the high viscosity donor material, due to the modal instability of the higher-order vortex mode.

To fully understand the mechanism, a theoretical analysis based on Bohr's model, [31] including Plateau-Rayleigh instability, Kelvin-Helmholtz instability [32] to consider the effect of 'spin jet' speed, or high viscosity donor material is necessary.

## 4. Conclusions

We have discovered that the OAM of the optical vortex field can twist an ultrahigh viscosity donor film to form a micrometer-scale 'spin jet'. This unique phenomenon, manifesting the helical trajectory of the optical vortex, should provide an entirely new physical aspect of OAM optical fields and interaction with matter. Going beyond both conventional ink jet and laser induced forward mass transfer printing technologies, this phenomenon should also enable the ejection of micrometer-scale droplets of the donor material with any viscosity, even ultrahigh viscosity of 4 Pa·s (approximately 1000 times higher than that for water and 100 times higher than that for ink used for conventional ink jet printing).

Higher-order vortex modes with $\ell>2$ will further accelerate the spinning motion of 'spin jet' and further stabilize the straight flight of 'spin jet'. Thus, the donor dots will be patterned on a receiver substrate separated by an extremely long distance beyond centimeter-scale. Such higher-order OAM experiments will be performed as a future work by employing a spatial light modulator.

The formation and ejection of such micrometer-scale 'spin jet's based on optical vortex induced forward mass transfer technique can be potentially extended to entirely novel patterning techniques, which could enable the development of next generation printed photonic/electronic/spintronic circuits formed by viscous droplets containing functional nanoparticles, such as quantum dots, metallic particles and magnetic ferrite particles, with ultrahigh spatial resolution and extremely long flight distance. The OV-LIFT patterning technique can also explore a completely new needleless drug injection.

**Funding**

The authors acknowledge support in the form of KAKENHI Grants-in-Aid (JP 16H06507; JP 17K19070; JP 18H03884) from the Japan Society for the Promotion of Science (JSPS), and Japan Science and Technology Agency (JST) CREST Grant (JPMJCR1903).

**Disclosures**

The authors declare no conflicts of interest.

**References**

1. J. Bharathan and Y. Yang, "Polymer electroluminescent devices processed by inkjet printing: I. Polymer light-emitting logo," Appl. Phys. Lett. **72**(21), 2660–2662 (1998).
2. B. Chen, T. Cui, Y. Liu, and K. Varahramyan, "All-polymer RC filter circuits fabricated with inkjet printing technology," Solid. State. Electron. **47**(5), 841–847 (2003).
3. S. E. Burns, P. Cain, J. Mills, J. Wang, and H. Sirringhaus, "Inkjet Printing of Polymer Thin-Film Transistor Circuits," MRS Bull. **28**(11), 829–834 (2003).
4. I. H. Choi and J. Kim, "A pneumatically driven inkjet printing system for highly viscous microdroplet formation," Micro Nano Syst. Lett. **4**(1), 4 (2016).
5. P. Serra and A. Piqué, "Laser-Induced Forward Transfer: Fundamentals and Applications," Adv. Mater. Technol. **4**(1), 1800099 (2019).
6. J. M. Fernández-Pradas, P. Sopeña, S. González-Torres, J. Arrese, A. Cirera, and P. Serra, "Laser-induced forward transfer for printed electronics applications," Appl. Phys. A Mater. Sci. Process. **124**(2), 214 (2018).
7. E. C. P. Smits, A. Walter, D. M. De Leeuw, and K. Asadi, "Laser induced forward transfer of graphene," Appl. Phys. Lett. **111**(17), 173101 (2017).
8. L. Allen, M. W. Beijersbergen, R. J. C. Spreeuw, and J. P. Woerdman, "Orbital angular momentum of light and the transformation of Laguerre-Gaussian laser modes," Phys. Rev. A **45**(11), 8185–8189 (1992).
9. J. Leach, M. J. Padgett, S. M. Barnett, S. Franke-Arnold, and J. Courtial, "Measuring the Orbital Angular Momentum of a Single Photon," Phys. Rev. Lett. **88**(25), 257901 (2002).
10. G. Molina-Terriza, J. P. Torres, and L. Torner, "Twisted photons," Nat. Phys. **3**(5), 305–310 (2007).
11. M. Padgett and R. Bowman, "Tweezers with a twist," Nat. Photonics **5**(6), 343–348 (2011).
12. S. M. Barnett, R. P. Cameron, S. M. Barnett, L. Allen, R. P. Cameron, S. Y. Buhmann, D. T. Butcher, S. Scheel, S. M. Barnett, and R. P. Cameron, "On the natures of the spin and orbital parts of optical angular momentum Energy conservation and the constitutive relations in chiral and non-reciprocal media," J. Opt. **18**(6), 064004 (2016).
13. M. Soskin, S. V. Boriskina, Y. Chong, M. R. Dennis, and A. Desyatnikov, "Singular optics and topological photonics," J. Opt. **19**(1), 010401 (2017).
14. M. J. Padgett, "Orbital angular momentum 25 years on [Invited]," Opt. Express **25**(10), 11265–11274 (2017).
15. A. T. O'Neil, I. MacVicar, L. Allen, and M. J. Padgett, "Intrinsic and Extrinsic Nature of the Orbital Angular Momentum of a Light Beam," Phys. Rev. Lett. **88**(5), 053601 (2002).
16. T. Omatsu, K. Chujo, K. Miyamoto, M. Okida, K. Nakamura, N. Aoki, and R. Morita, "Metal microneedle fabrication using twisted light with spin," Opt. Express **18**(17), 17967–17973 (2010).
17. K. Toyoda, K. Miyamoto, N. Aoki, R. Morita, and T. Omatsu, "Using optical vortex to control the chirality of twisted metal nanostructures," Nano Lett. **12**(7), 3645–3649 (2012).

18. K. Toyoda, F. Takahashi, S. Takizawa, Y. Tokizane, K. Miyamoto, R. Morita, and T. Omatsu, "Transfer of light helicity to nanostructures," Phys. Rev. Lett. **110**(14), 143603 (2013).
19. M. Watabe, G. Juman, K. Miyamoto, and T. Omatsu, "Light induced conch-shaped relief in an azo-polymer film," Sci. Rep. **4**, 4281 (2014).
20. F. Takahashi, S. Takizawa, H. Hidai, K. Miyamoto, R. Morita, and T. Omatsu, "Optical vortex pulse illumination to create chiral monocrystalline silicon nanostructures," Phys. Status Solidi Appl. Mater. Sci. **213**(4), 1063–1068 (2016).
21. F. Takahashi, K. Miyamoto, H. Hidai, K. Yamane, R. Morita, and T. Omatsu, "Picosecond optical vortex pulse illumination forms a monocrystalline silicon needle," Sci. Rep. **6**, 21738 (2016).
22. M. W. Beijersbergen, R. P. C. Coerwinkel, M. Kristensen, and J. P. Woerdman, "Helical-Wave-Front Laser-Beams Produced With a Spiral Phaseplate," Opt. Commun. **112**(5–6), 321–327 (1994).
23. D. B. Bogy, "Drop Formation in a Circular Liquid Jet," Annu. Rev. Fluid Mech. **11**, 207–228 (1979).
24. R. Mead-Hunter, A. J. C. King, and B. J. Mullins, "Plateau rayleigh instability simulation," Langmuir **28**(17), 6731–6735 (2012).
25. S. Kawano, "Molecular dynamics of rupture phenomena in a liquid thread," Phys. Rev. E **58**(4), 4468–4472 (1998).
26. C. J. Knight, "Theoretical Modeling of Rapid Surface Vaporization with Back Pressure," AIAA J. **17**(5), 519–523 (1979).
27. J. Zhou, H.-L. Tsai, and P.-C. Wang, "Transport Phenomena and Keyhole Dynamics during Pulsed Laser Welding," J. Heat Transfer **128**(7), 680–690 (2006).
28. R. Nagura, T. Tsujimura, T. Tsuji, K. Doi, and S. Kawano, "Coarse-grained particle dynamics along helical orbit by an optical vortex irradiated in photocurable resins," OSA Contin. **2**(2), 400–415 (2019).
29. P. R. N. Childs, *Rotating Flow*, 1st Edition (Elsevier, New York, 2010).
30. E. W. Lemmon and R. T. Jacobsen, "Viscosity and Thermal Conductivity Equations for Nitrogen, Oxygen, Argon, and Air," Int. J. Thermophys. **25**(1), 21–69 (2003).
31. N. Bohr, "Determination of the Surface-Tension of Water by the Method of Jet Vibration," Philos. Trans. R. Soc. London. Ser. A, Contain. Pap. a Math. or Phys. Character **209**, 281–317 (n.d.).
32. S. Kawano, H. Hashimoto, H. Togari, A. Ihara, T. Suzuki, and T. Harada, "Deformation and breakup of annular liquid sheet in a gas stream," At. Sprays **7**(4), 359–374 (1997).